\documentclass[10pt,twocolumn,aps,prb]{revtex4-1}

\usepackage{amsfonts}
\usepackage{amsmath}
\usepackage{amssymb}
\usepackage{bm}
\usepackage{graphicx}
\usepackage[utf8]{inputenc}
\usepackage{textcomp}

\providecommand{\vect}[1]{\boldsymbol{#1}}
\DeclareMathOperator{\IM}{Im}
\DeclareMathOperator{\RE}{Re}
\DeclareMathOperator{\trace}{tr}
\DeclareMathOperator{\SIGN}{sign}

\begin{document}

%%%%%%%%%%%%%%%%%%%%%%%%%%%%%%%%%%%%%%%%%%%%%%%%%%%%%%%%%%%%%%%%%%%%%%%%%%%%%%%%
% MAIN TEXT                                                                    %
%%%%%%%%%%%%%%%%%%%%%%%%%%%%%%%%%%%%%%%%%%%%%%%%%%%%%%%%%%%%%%%%%%%%%%%%%%%%%%%%

\title{%
Interaction Correction to the Magneto-Electric Polarizability \\
of $Z_{2}$ Topological Insulators%
}
\author{Karin Everschor-Sitte}
\author{Matthias Sitte}
\author{Allan H. MacDonald}
\affiliation{%
Department of Physics,
The University of Texas at Austin,
Austin, TX 78712, USA%
}
\date{\today}

\begin{abstract}
When time-reversal symmetry is weakly broken and interactions are neglected,
the surface of a $Z_{2}$ topological insulator supports a half-quantized Hall
conductivity $\sigma_{S} = e^{2}/(2h)$.  A surface Hall conductivity in an
insulator is equivalent to a bulk magneto-electric polarizability,
\textit{i.e.}\ to a magnetic field dependent charge polarization.  By
performing an explicit calculation for the case in which the surface is
approximated by a two-dimensional massive Dirac model and time-reversal
symmetry is broken by weak ferromagnetism in the bulk, we demonstrate that
there is a non-universal interaction correction to $\sigma_{S}$.  For thin
films interaction corrections to the top and bottom surface Hall conductivities
cancel, however, implying that there is no correction to the quantized
anomalous Hall effect in magnetically doped topological insulators.
\end{abstract}

\maketitle

%%%%%%%%%%%%%%%%%%%%%%%%%%%%%%%%%%%%%%%%%%%%%%%%%%%%%%%%%%%%%%%%%%%%%%%%%%%%%%%%

\section{Introduction}
\label{sec:Introduction}

The quantum Hall effect\cite{Klitzing1980} stands alone among transport
phenomena because it is characterized by a non-zero transport coefficient whose
value is universal, dependent only on fundamental constants of nature and not
at all on crystal imperfections and other peculiarities of individual samples.
The accuracy of the quantum Hall effect is now established to better than eight
figures\cite{Tzalenchuk2010} and has no established limitation.  This
surprising property can be traced to its identification with a topological
index\cite{Thouless1982, Haldane1988, Ryu2010} of electronic structure, one
that can be non-trivial only in systems with broken time-reversal symmetry. For
many years quantum Hall states endured as the only known example of
topologically non-trivial electronic structure.  In recent years, however, the
topological classification\cite{Kitaev2009, Ryu2010} of electronic states has
broadened considerably. The $Z_{2}$ classification\cite{Kane2005, Bernevig2006,
Fu2006, Koenig2007, Fu2007} of what are seemingly the most innocent of
states---time-reversal invariant insulators---has particularly broad
experimental implications.  Only in the original quantum Hall case, however, is
the topological index a readily measured macroscopic observable.

Non-trivial electronic topology is most commonly revealed by the presence of
protected boundary states at surfaces and heterojunctions.\cite{Qi2011,
Hasan2010}  The topological character of a three-dimensional insulator, for
example, can be revealed by examining its surface states\cite{Hsieh2008} to
determine whether the number of Dirac points (linear band crossings) is even or
odd.  The observable that is most closely related to the non-trivial $Z_{2}$
topological index of time-reversal invariant insulators is its magneto-electric
polarizability,\cite{Qi2008, Essin2009, Essin2010, Malashevich2010} or
equivalently its surface-state Hall conductivity.  Because a finite Hall
conductivity requires broken time-reversal symmetry, the association of
magneto-electric polarizability with a time-reversal invariant state is
puzzling.  The accepted resolution\cite{Coh2011} of this conundrum, briefly, is
that the bulk magneto-electric polarizability is observable only when
time-reversal invariance is weakly broken at the surface and the Fermi level
lies in the resulting surface-state gap.  When these conditions are satisfied,
it is commonly argued that the surface Hall conductivity of a non-interacting
$Z_{2}$ topological insulator (TI) must be quantized at a half-odd-integer
multiple of $e^{2}/h$ because i) it must change sign under time reversal, and
ii) it can change only by integer multiples of $e^{2}/h$ under time-reversal or
under any other change in the Hamiltonian.  This magneto-electric response of a
TI has been referred to as its Chern-Simons polarizability.  In this article we
show that, in contrast to the case of the quantum Hall effect, weak
interactions quite generally yield a correction to this observable.

\begin{figure}[b]
\centering
\includegraphics[width=.4\textwidth]{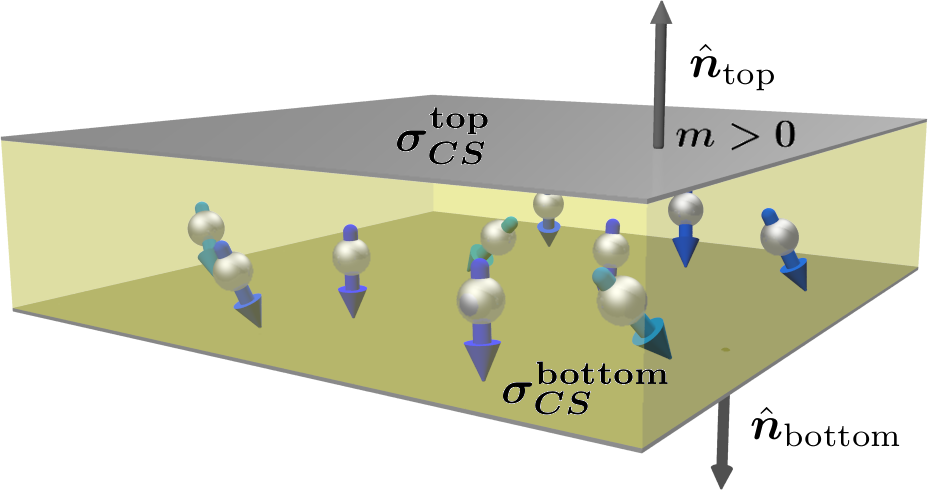}
\caption{%
A diluted-moment topological-insulator ferromagnet containing local-moment
spins that order, breaking time-reversal symmetry and coupling to its surface
Dirac cones.  We show that interactions between surface-state quasi-particles
and fluctuations of the magnetic condensate are responsible for corrections of
opposite sign to the top and bottom surface half-quantized Hall
conductivities.%
}
\label{fig:SketchDopedTI}
\end{figure}

Our conclusions are based on an explicit calculation for the case of a TI
surface with a single Dirac cone, and time-reversal symmetry that is broken by
weak bulk ferromagnetism (see Fig.~\ref{fig:SketchDopedTI}). The model we
consider provides a good description of the thin-film diluted-moment
ferromagnets based on (Bi,Sb)$_2$Te$_3$ TIs in which the quantized anomalous
Hall effect (QAHE)\cite{Chang2013, Kou2014, Checkelsky2014, Bestwick2014} has
recently been observed.  Chromium or vanadium doping in these materials
introduces local moments that order at low temperatures, breaking time-reversal
symmetry and opening a gap in the surface-state spectrum.  The
discovery\cite{Chang2013} of a QAHE in this material was inspired by earlier
theoretical work\cite{Yu2010} which predicted that thin films of the
tetradymite semiconductors Bi$_{2}$Te$_{3}$, Bi$_{2}$Se$_{3}$, and
Sb$_{2}$Te$_{3}$ would reveal a quantized Hall effect when doped with
transition metal elements.

The Hall conductivity on both top and bottom surfaces of a diluted-moment TI
ferromagnet is expected to be half-quantized,\cite{Qi2008, Essin2009, Tse2010a}
provided\cite{Sitte2012} that time-reversal-symmetry breaking energy scales are
small compared to the bulk energy gap. When electronic properties of the system
are evaluated using mean-field theory, this expectation is corroborated in the
small surface-state-gap limit by calculations based on a Dirac model with an
energy gap due to exchange interactions between surface-state quasi-particles
and the bulk magnetic condensate.\cite{Qi2011, Hasan2010}  We show below that
the surface Hall effect is no longer exactly half-quantized when interactions
between surface-state quasi-particles and quantum fluctuations of the bulk
magnetization, described as magnons, are included.  The total Hall effect
obtained by summing over the top and bottom surfaces of a thin film remains
quantized however, in agreement with experiment.

%%%%%%%%%%%%%%%%%%%%%%%%%%%%%%%%%%%%%%%%%%%%%%%%%%%%%%%%%%%%%%%%%%%%%%%%%%%%%%%%

\section{Surface-State Hamiltonian}
\label{sec:SurfaceStateHamiltonian}

We consider two-dimensional (2D) surface-state model Hamiltonians with a single
Dirac cone, exchange interactions, and spin-dependent disorder or interaction
terms:
\begin{equation}
H = H_{\mathrm{qp}} + H_{\mathrm{pert}},
\end{equation}
where $H_{\mathrm{qp}}$ is a mean-field-theory quasi-particle Hamiltonian for a
gapped Dirac system, and $H_{\mathrm{pert}}$ is a perturbation.  The mean-field
Hamiltonian can quite generally be expressed in the form
\begin{equation}
\label{eq:BareElHamiltonian}
H_{\mathrm{qp}} =\sum_{\vect{k}} \Psi_{\vect{k}}^{\dagger}
\mathcal{H}_{\mathrm{qp}}(\vect{k}) \Psi_{\vect{k}},
\end{equation}
where $\Psi_{\vect{k}}$ is an annihilation operator spinor, and
$\mathcal{H}_{\mathrm{qp}}(\vect{k})$ is expanded in a Pauli matrix basis:
\begin{equation}
\label{eq:BareElHamiltonianBloch}
\mathcal{H}_{\mathrm{qp}}(\vect{k}) = d_0(\vect{k}) \sigma_{0} +
\vect{d}(\vect{k}) \cdot \vect{\sigma}.
\end{equation}
This Hamiltonian has a gap separating low-energy valence-band surface states,
which are occupied in the case of interest, from high-energy conduction-band
surface states:
\begin{equation}
\xi_{\pm}(\vect{k}) = d_0(\vect{k}) \pm |\vect{d}(\vect{k})|.
\end{equation}
When the surface-state Hamiltonian is time-reversal invariant, $\vect{d}$ and
hence the gap must vanish at $\vect{k} = 0$.  In order to clearly explain the
origin of the surface-state Hall conductivity correction, we specialize below
to the case of the 2D massive Dirac model which is simplified by isotropic
energy bands:
\begin{equation}
\label{eq:MassiveDiracHamiltonian}
\mathcal{H}^{\mathrm{MD}}_{\mathrm{qp}}(\vect{k}) = \hbar v \hat{\vect{z}}
\cdot (\vect{k} \times \vect{\sigma}) \pm \hbar m \sigma_{z} \equiv
\vect{d}^{\mathrm{MD}}_{\pm}(\vect{k}) \cdot \vect{\sigma},
\end{equation}
where we have chosen the zero of energy at the Dirac point, $v$ is the Fermi
velocity of the surface-state Dirac fermions, $\Delta = 2 \hbar |m|$ is the
surface-state gap, and the sign in Eq.~(\ref{eq:MassiveDiracHamiltonian})
depends on the direction of the thin-film magnetization relative to the surface
normal.  The $\sigma_{z}$ term in this Hamiltonian is the mean-field exchange
interaction between the surface-state spins and perpendicular anisotropy bulk
magnetization.

We describe our Hall conductivity calculation in detail for the case in which
the surface normal and the exchange field on the surface are parallel and in
the $\hat{z}$ direction. This choice corresponds to spin-$\downarrow$ occupied
surface states and, if the interaction between the surface state quasi-particle
and the bulk magnetization is ferromagnetic, to a spin-$\downarrow$ bulk spin
orientation. The gapped surface-state conduction- and valence-band energies are
given by:
\begin{equation}
\xi_{\pm}^{\mathrm{MD}}(\vect{k}) = \pm \hbar \sqrt{v^{2} |\vect{k}|^{2} +
m^{2}}.
\end{equation}

We distinguish two types of perturbative corrections to the massive Dirac
model:  i) static perturbations in which the Hamiltonian is changed but the
Hilbert space is not, and ii) dynamic perturbations in which the surface-state
quasi-particle are coupled to external bosonic degrees of freedom like phonons
or magnons.  In the first case, we consider the Hamiltonian
$\mathcal{H}_{\mathrm{pert}}^{\mathrm{st}} = g_{0} \sigma_{0} + \vect{g} \cdot
\vect{\sigma}$, where $g_{0}$ and $\vect{g}$ are charge and spin disorder
potentials that depend randomly on position.  Since in this article our goal is
simply to establish that the interaction corrections to the Hall conductivity
do not vanish, we calculate corrections only to leading order in perturbation
theory.  Because the leading order response can be written as a sum over
contributions from different Fourier components $\vect{p}$ of $g_{0}$ and
$\vect{g}$, we can consider one component at a time.  It is therefore
sufficient to assume that these functions vary sinusoidally with position with
arbitrary wavevector $\vect{p}$.

In the dynamic perturbation case, $H^{\mathrm{dy}}_{\mathrm{pert}} =
H_{\mathrm{b}} + H_{\mathrm{qp-b}}$, we add to the Hamiltonian both a bare
boson contribution $H_{\mathrm{b}}$ and an interaction $H_{\mathrm{qp-b}}$
between quasi-particles and bosons:
\begin{subequations}
\begin{align}
H_{\mathrm{b}} &= \sum_{\vect{p}} \hbar \omega_{\vect{p}}
a_{\vect{p}}^{\dagger} a_{\vect{p}}, \\
H_{\mathrm{qp-b}} &= A^{-1/2} \sum_{\vect{k}} (\Psi_{\vect{k} -
\vect{p}}^{\dagger} a_{\vect{p}}^{\dagger} \mathcal{M} \Psi_{\vect{k}} +
\text{h.c.}).
\end{align}
\end{subequations}
Here, $a_{\vect{p}}^{\dagger}$ ($a_{\vect{p}}$) creates (annhilates) bosons
with momentum $\vect{p}$, $\omega_{\vect{p}}$ specifies the boson dispersion,
$A$ is the surface area, and $\mathcal{M}$ is a electron-boson interaction
coupling matrix which can be spin-dependent.  In the zero temperature limit, we
can, in calculating the leading-order electron-boson interaction correction,
truncate the boson Hilbert space both to a single boson momentum $\vect{p}$ and
to the $n = 0$ and $n = 1$ occupation numbers.  These simplifications allow the
dressed eigenstates to be obtained by diagonalizing $4 \times 4$ matrices for
each $\vect{k}$.

Because the exchange interaction between a magnetic quasi-particle and a
ferromagnetic condensate is (at least approximately) invariant under
simultaneous rotation of the magnetic order parameter and the quasi-particle
spin, magnon creation (which raises spin for the $\downarrow$ condensate spin
direction considered here) is accompanied by quasi-particle spin-flip from
$\uparrow$ to $\downarrow$ and magnon annihilation by quasi-particle spin-flip
from $\downarrow$ to $\uparrow$.  We therefore write $\mathcal{M}_{\mathrm{sw}}
= \gamma_{\mathrm{sw}} (\sigma_{x} - i \sigma_{y})/2$. We show below that this
interaction vertex implies a correction to the surface Hall conductivity.

%%%%%%%%%%%%%%%%%%%%%%%%%%%%%%%%%%%%%%%%%%%%%%%%%%%%%%%%%%%%%%%%%%%%%%%%%%%%%%%%

\section{Magneto-Electric Polarizability}
\label{sec:MagnetoElectricPolarizability}

Using linear-response theory (see Sec.~\ref{supp:DerivationHallConductivity} of
the supplemental material), the surface-state Hall conductivity can be
expressed in terms of current-operator matrix elements between
momentum-dependent ground $|{0}\rangle$ and excited states $|{n}\rangle$:
\begin{subequations}
\label{eq:sigmaxyGeneralExpression}
\begin{align}
\sigma_{xy}
\label{eq:sigmaxyGeneralExpressionA}
&= -\frac{ \hbar }{2 \pi^2} \int_{\mathrm{DP}} d^{2}k \sum_{n \neq 0}
\frac{\IM(\langle{0}| j_{x} |{n}\rangle \langle {n}| j_{y} |{0}\rangle)}{(E_{n}
- E_{0})^{2}/\hbar^2} \\
\label{eq:sigmaxyGeneralExpressionB}
&= \frac{e^{2}}{2 \pi h} \int_{\mathrm{DP}} d^{2} k\ \Omega_{xy}(\vect{k}) \\
\label{eq:sigmaxyGeneralExpressionC}
&= \frac{e^{2}}{2 \pi h} \oint_{\partial \mathrm{DP}} d\vect{k} \cdot
\vect{A}(\vect{k}).
\end{align}
\end{subequations}
In Eq.~\eqref{eq:sigmaxyGeneralExpression} the integrals over momentum are
taken over the Dirac point region $\mathrm{DP}$, bounded by $\partial
\mathrm{DP}$, defined as the region in which the surface states lie inside the
bulk gap. Eqs.~(\ref{eq:sigmaxyGeneralExpressionB}) and
(\ref{eq:sigmaxyGeneralExpressionC}) rely on the observation that the continuum
model current operator expression, $j_{\mu} = -(e/\hbar) (\partial H/\partial
k_{\mu})$, remains valid when electron-boson coupling is included. When the
boson momentum is restricted to $\vect{p}$ and the boson Hilbert space is
truncated to $n=0,1$, the eigenstates in
Eq.~\eqref{eq:sigmaxyGeneralExpression} are linear combinations of $n = 0$ band
electron states with momentum $\vect{k}$, and $n = 1$ band states with momentum
$\vect{k-p}$. The Berry curvature\cite{Xiao2010} is given by:
\begin{align}
\label{eq:sigmaxyBerryCurvature}
\Omega_{xy}(\vect{k})
&= i \sum_{n \neq 0} \frac{\langle{0}| \frac{\partial H}{\partial k_{x}}
|{n}\rangle \langle {n}| \frac{\partial H}{\partial k_{y}} |{0}\rangle - (x
\leftrightarrow y)}{(E_{n} - E_{0})^{2}} \nonumber \\
&= \partial_{k_{x}} A_{y}(\vect{k}) - \partial_{k_{y}} A_{x}(\vect{k}),
\end{align}
where the Berry connection $A_{\mu}(\vect{k}) = i \langle {0}|
\partial_{k_{\mu}} |{0}\rangle$.  When applying
Eq.~\eqref{eq:sigmaxyGeneralExpressionC} we must choose a gauge in which the
ground state is a smooth function of wavevector inside the region
$\mathrm{DP}$.

In the absence of interactions and disorder (\textit{i.e.}\ for
$H_{\mathrm{pert}} = 0$), Eq.~\eqref{eq:sigmaxyGeneralExpressionA} reduces to
\begin{equation}
\label{eq:UnperturbedSigmaXY}
\sigma_{xy} = -\frac{\hbar}{2 \pi^2 } \int_{\mathrm{DP}} d^{2} k\
\frac{\IM(\langle{0}| j_{x} |{1}\rangle \langle{1}| j_{y} |{0}\rangle)}{(E_{1}
- E_{0})^{2}},
\end{equation}
where $|{0}\rangle$ now represents a valence band and $|{1}\rangle$ a
conduction band single-particle state.  Performing the wavevector integration
recovers the half-integer QAHE obtained in independent-particle
theories:\cite{Redlich1984, Qi2008}
\begin{equation}
\sigma_{xy} = \SIGN(\mathcal{V}) \SIGN(m) \frac{e^{2}}{2h},
\end{equation}
where by $\mathcal{V}$ we denote the sense of the vorticity of the
momentum-space valence-band-spinor texture in the absence of a gap.  The same
result for the Hall conductivity can be obtained by using the Berry connection
expression.   For the massive Dirac model the line integral in
Eq.~\eqref{eq:sigmaxyGeneralExpressionC} is around a circle with radius
$\Lambda$ such that $v \Lambda \gg m$.
Eq.~\eqref{eq:sigmaxyGeneralExpressionC} then simplifies to
\begin{equation}
\label{eq:sigmaxyOurModelA}
\sigma_{xy} = \frac{e^{2}}{2 \pi h} \int_{0}^{2\pi} d\phi\, i\, \langle{0}|
\frac{\partial}{\partial \phi} |{0}\rangle\vert_{k=\Lambda}.
\end{equation}
where $\phi$ is the momentum orientation angle.  We use this expression below
to calculate the correction to the surface state Hall conductivity when
electron-magnon interactions are included.

As explained previously, the half-quantized surface state Hall conductivity is
expected to be invariant under weak perturbations.  In
Sec.~\ref{supp:QuenchedSpinWave} of the supplemental material we demonstrate
explicitly that this expectation is confirmed when the massive Dirac
single-particle Hamiltonian is perturbed by a weak spin-dependent disorder
term. However, as we now show, corrections are finite when the Dirac
surface-state quasi-particle interact with quantum fluctuations of the ordered
state responsible for time-reversal symmetry breaking.

\begin{figure*}[tb]
\includegraphics[width=0.95\textwidth]{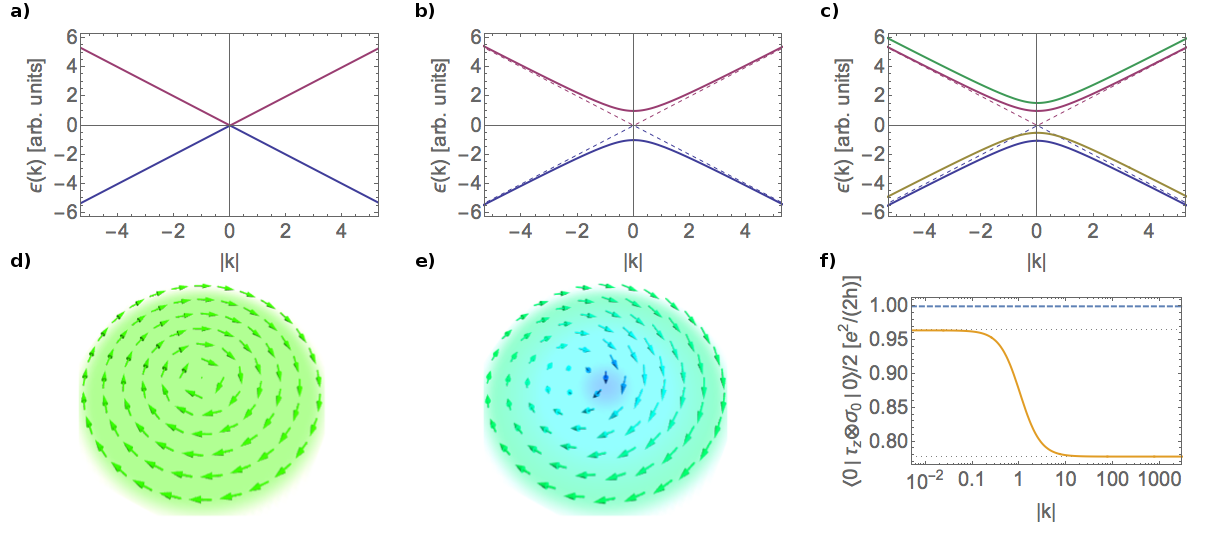}
\caption{%
(Color online) Band structure for \textbf{a)} a pure $(\hbar v = 1$, $m/v = 0)$
Dirac model, \textbf{b)} a massive ($m/v = 1$) Dirac model, and \textbf{c)} a
massive Dirac model interacting with momentum $\vect{p} = 0$ magnons restricted
to occupation numbers 0 and 1 ($A^{-1/2} \Omega/v = 1/3$, $\omega/v = 1/2$).
Panel \textbf{d)} shows the momentum space spin texture of the ground state of
the pure Dirac model in which spins projections lie in the $xy$ plane and
rotate along with the momentum direction.  Panel \textbf{e)} shows the spin
texture of the massive Dirac model with a momentum-space vortex centered at
$\vect{k} = 0$. The spin is in the $-\hat{z}$ direction at $\vect{k} = 0$.
(The color code denotes the $z$ component of the spins.) Panel \textbf{f)}
shows the result of Eq.~\eqref{eq:sigmaxyOurModelA} in units of $e^{2}/(2h)$ as
a function of $|k|$ in the non-interacting and the
electron-spinwave-interacting 2D massive Dirac model.  For large $|k|$ the
interacting model does not converge to the quantized value of $e^{2}/(2h)$ but
obtains a correction given by $[-(\Omega/\omega)^{2}/2] \times e^{2}/(2h)$.%
}
\label{fig:MainFigure}
\end{figure*}

The origin of the interaction effect is schematically summarized in
Fig.~\ref{fig:MainFigure} where we illustrate (panels \textbf{a}--\textbf{c})
the surface-state band structure of the massless  Dirac model, the massive
Dirac model, and the Dirac model coupled to a bosonic mode.  The band
eigenstates can be viewed (panels \textbf{d} and \textbf{e}) as
momentum-dependent spin-$1/2$ coherent states.  When electron-magnon coupling
is neglected the massive Dirac model spin has spin-$\downarrow$ orientation at
the Dirac point $\vect{k} = 0$, and an in-plane orientation at large $|k|$ with
a finite vorticity, forming a meron.  The $\vect{k} = 0$ spin orientation fixes
the gauge choice for the unperturbed spin-coherent states.  Because of the
large splitting between conduction- and valence-band states at large $|k|$ used
to evaluate the Berry connection, electron-magnon scattering coherently mixes
primarily $n = 0$ and $n = 1$ magnon states, leaving the electronic state in
the valence band.  The Hall conductivity correction is due in part to the
reduced weight of the $n = 0$ valence-band state responsible for the
non-interacting Hall effect, and in part due to the momentum-orientation
coherence between $n = 0$ and $n = 1$ states which changes the sign of the $n =
1$ Berry connection contribution. In panel \textbf{f} of
Fig.~\ref{fig:MainFigure} we plot the Berry connection integral of
Eq.~\eqref{eq:sigmaxyOurModelA}, calculated as a function of $|k|$ both
neglecting and including electron-magnon interactions.  For large $|k|$ the
interacting model does not converge to the quantized value of $1/2$ but obtains
an interaction correction.  The calculation is described in greater detail
below.

At leading order in perturbation theory, corrections are obtained by summing
over contributions from distinct boson modes, and the boson Hilbert space can
be truncated to occupation numbers 0 and 1.  To bring out the physics of the
interaction correction as simply as possible we focus first on the contribution
from interactions between surface-state quasiparticles and a boson mode with 2D
momentum $\vect{p} = 0$.  This simplification leads to a Hilbert space in which
four possible states are associated with each crystal momentum, valence- and
conduction-band states with and without a boson present.  The many-body
Hamiltonian is then diagonal in crystal momentum, and each $4 \times 4$ block
has the form
\begin{equation}
\mathcal{H}^{n=1} =
\begin{pmatrix}
  \mathcal{H}_{\mathrm{qp}} & \mathcal{M}\\
  \mathcal{M}^{\dagger} & \mathcal{H}_{\mathrm{qp}} + \hbar \omega
\end{pmatrix}.
\end{equation}
For electron-magnon interactions the spin-dependent quasi-particle-boson
interaction matrix\cite{footnote1}
\begin{equation}
\mathcal{M} = \hbar \, \Omega
\begin{pmatrix}
  \mathcal{M}_{11} & \mathcal{M}_{12} \\
  \mathcal{M}_{21} & \mathcal{M}_{22}
\end{pmatrix}
\end{equation}
has only one non-zero element since magnon creation is accompanied by spin-flip
from $\uparrow$ to $\downarrow$:
\begin{equation}
\Omega \mathcal{M}_{21} = \frac{m}{2 \sqrt{M_0}}
\end{equation}
where $m$ is the quasi-particle mass, and $M_{0}$ is spin per unit area of the
thin film.

To calculate the Hall conductivity correction we separate $\mathcal{H}^{n=1}$
into $\mathcal{H}^{n=1}_{0}$ and $\mathcal{H}^{n=1}_{\mathrm{pert}}$ with
\begin{equation}
\mathcal{H}^{n=1}_{0} =
\begin{pmatrix}
  \mathcal{H}_{\mathrm{qp}} & 0\\
  0 & \mathcal{H}_{\mathrm{qp}} + \hbar \omega
\end{pmatrix},
\mathcal{H}^{n=1}_{\mathrm{pert}} =
\begin{pmatrix}
  0 & \mathcal{M} \\
  \mathcal{M}^{\dagger} & 0
\end{pmatrix}.
\end{equation}
For $m > 0$, the unperturbed ground state at $\vect{k} = 0$ is a
spin-$\downarrow$ state.  At finite $\vect{k}$ the unperturbed ground state is
a spin-coherent state with a finite in-plane component with orientation $\chi =
\phi + \pi/2$.  In order to use the Berry phase formula for the Hall
conductivity we must choose the gauge in which the phase factor $\exp(-i \chi)$
appears in the spin-$\uparrow$ component of the unperturbed ground state
spinor. The correction to the ground state due to interactions with magnons can
then be calculated using first-order perturbation theory.  At large wavevectors
we can ignore mixing between conduction- and valence-band states because of the
large $v \Lambda$ energy denominator.  In this way we find that on $\partial
\mathrm{DP}$:
\begin{equation}
\label{eq:pertgroundstate}
|{0}\rangle \approx |{n=0}\rangle \otimes |{v}\rangle - \frac{\Omega M_{21}
\exp(i \chi)}{2 \omega} |{n=1}\rangle \otimes |{v}\rangle,
\end{equation}
where
\begin{equation}
|v\rangle = \frac{1}{\sqrt{2}} (\exp(-i \chi), 1)
\end{equation}
is the unperturbed valence band state on $\partial \mathrm{DP}$.  It then
follows from the Berry connection formula for the Hall conductivity that
\begin{equation}
\label{eq:BasicResult}
\sigma_{xy} \approx \frac{e^{2}}{2h} \SIGN{(\mathcal{V})} \biggl[ \SIGN (m) -
\frac{1}{2} \biggl( \frac{\Omega}{\omega} \biggr)^{2} |\mathcal{M}_{21}|^{2}
\biggr].
\end{equation}
In Eq.~\eqref{eq:BasicResult} we have generalized to the cases in which the
surface-state Dirac model is altered by changing the sign of the mass $m$
and/or the vorticity of momentum-space spin texture. ($\chi =
\SIGN{(\mathcal{V})}(\phi + \pi/2$).)

Because the valence-band states on $\partial \mathrm{DP}$ vary with momentum on
the scale of $\Lambda$, the magnon-mode Hall conductivity correction
calculation at finite $\vect{p}$ is unchanged relative to $\vect{p} = 0$
provided that the momentum magnitude $|\vect{p}|$ that is much smaller than
$\Lambda$. An expression for the Hall conductivity correction valid for
arbitrary electron-boson interaction vertex and arbitrary surface-state
band-structure model requires a lengthy and detailed calculation, and is
provided in Sec.~\ref{supp:GeneralInteractionVertex} of the supplemental
material.  For a diluted-moment magnetically ordered TI thin film, the
quasi-particle mass and the quasi-particle vorticity are both opposite in sign
on top and bottom surfaces.  It follows that, although the Hall conductivities
of the top and bottom surfaces both have corrections, they differ in sign.

%%%%%%%%%%%%%%%%%%%%%%%%%%%%%%%%%%%%%%%%%%%%%%%%%%%%%%%%%%%%%%%%%%%%%%%%%%%%%%%%

\section{Discussion}

In the previous section we calculated the contribution of a single magnon mode
to the Hall conductivity interaction correction, which is inversely
proportional to the surface area of the system.  The correction to the Hall
conductivity varies slowly with magnon momentum $\vect{p}$ provided $\vect{p}$
is close to the Dirac point.  Summing over magnons with momenta inside
$\mathrm{DP}$ we predict an overall correction proportional to
$(A_{\mathrm{DP}}/M_{0}) (m/\omega)^{2}$, where $A_{\mathrm{DP}}$ is the area
in momentum space of the Dirac point region $\mathrm{DP}$.  Since the gap in
the magnon spectrum, due either to weak external fields used to saturate the
magnetization or to the perpendicular magnetic anisotropy of magnetically doped
TI thin films, is typically smaller than the gap produced in the surface-state
quasi-particle spectrum, the interaction correction can be large even when $m
\ll v \Lambda $.  A large interaction correction to the magneto-electric
coefficients of TI thin films is present even when time-reversal symmetry
breaking is weak when measured by the size of the surface-state gap it
produces.  This result, which may seem surprising, is in fact natural because
of the strong spin-orbit coupling inevitably present in TIs.  A magnetic order
parameter in a magnetically doped TI will never be a good quantum number.
Quantum fluctuations of the magnetic condensate interact with surface-state
quasi-particles and cause the system's broken time-reversal symmetry to be
manifested even in quasi-particles that are far from the Dirac point.

A TI differs from an ordinary insulator mainly via its protected surface
states, and these complicate\cite{Tse2010b, Morimoto2015} the task of measuring
the magneto-electric effects discussed here.  In particular, electrical
measurements of a magnetic field dependent film polarization are not possible
when the system has a non-zero total Hall conductivity, because this is
necessarily associated with edge states which are localized on side walls and
short the top and bottom  surfaces of the film.  As recently discussed in
Ref.~\citenum{Morimoto2015}, however, electrical measurements should be
feasible when the top half of the thin film is doped with Cr ions and the
bottom half with Mn ions.  These atoms have exchange interactions with
surface-state electrons that have opposite sign.  When they are aligned by a
weak magnetic field, the sign of the exchange effective field on top and bottom
surface Dirac cones is opposite.\cite{Checkelsky2012, Chang2013,
Checkelsky2014}  In terms of the massive Dirac models we have studied in this
paper, this circumstance implies that there are no side wall states and that
while the signs of the momentum-space vorticities on the top and bottom
surfaces are opposite, the masses have the same sign.  Because the total Hall
conductivity is zero in this case, there should be an energy range over which
there are no side wall states.  The individual surface Hall conductivities are
non-zero however, and they can be measured electrically by detecting current
flow between top and bottom surfaces as magnetic field strength is varied.  We
predict that this measurement will identify an interaction correction to the
surface state Hall conductivity.  Similar interaction corrections which
contribute to the valley Hall effect but cancel out in the total anomalous Hall
effect occur in honeycomb lattice Dirac systems\cite{Haldane1988, Xiao2007}
when the electron-boson interaction is sublattice dependent.

%%%%%%%%%%%%%%%%%%%%%%%%%%%%%%%%%%%%%%%%%%%%%%%%%%%%%%%%%%%%%%%%%%%%%%%%%%%%%%%%

\section{conclusions}

The surface Hall conductivity of an insulator is proportional to its
magneto-electric polarizability, \textit{i.e.}\ to the coefficient which
describes how the polarization of a film depends on magnetic field strength. By
explicitly evaluating the surface Hall conductivity of surface states described
by a massive Dirac model, we have shown that there is a non-universal
interaction correction to the quantized magneto-electric coefficient of thin
films formed from TIs.  Corrections to the top and bottom surface Hall
conductivities cancel, however, imply that there is no correction to the
quantized anomalous Hall effect in magnetically doped TIs.  The interaction
correction to the magneto-electric polarizability can be measured electrically
only when the total Hall conductivity of top and bottom surfaces is made to
vanish, for example by aligning local moments with opposite signs of exchange
coupling to the Dirac surface states.

%%%%%%%%%%%%%%%%%%%%%%%%%%%%%%%%%%%%%%%%%%%%%%%%%%%%%%%%%%%%%%%%%%%%%%%%%%%%%%%%

\section*{Acknowledgments}

This work was supported by the DOE Division of Materials Sciences and
Engineering under grant DE-FG03-02ER45958 and by the Welch foundation under
grant F1473.  KE and MS acknowledge financial support from DAAD.

%%%%%%%%%%%%%%%%%%%%%%%%%%%%%%%%%%%%%%%%%%%%%%%%%%%%%%%%%%%%%%%%%%%%%%%%%%%%%%%%

%%%%%%%%%%%%%%%%%%%%%%%%%%%%%%%%%%%%%%%%%%%%%%%%%%%%%%%%%%%%%%%%%%%%%%%%%%%%%%%%
% SUPPLEMENTAL MATERIAL                                                        %
%%%%%%%%%%%%%%%%%%%%%%%%%%%%%%%%%%%%%%%%%%%%%%%%%%%%%%%%%%%%%%%%%%%%%%%%%%%%%%%%

\onecolumngrid
\pagebreak

\begin{center}
\textbf{\large Supplemental Materials: Interaction Correction to the
Magneto-Electric Polarizability of $Z_{2}$ Topological Insulators}
\end{center}

%%%%
% Add a prefix "S" to all equations, figures, tables and reset the counters.
\setcounter{equation}{0}
\setcounter{figure}{0}
\setcounter{table}{0}
\setcounter{section}{0}
\setcounter{page}{1}
\makeatletter
\renewcommand{\theequation}{S\arabic{equation}}
\renewcommand{\thefigure}{S\arabic{figure}}
\renewcommand{\bibnumfmt}[1]{[S#1]}
\renewcommand{\citenumfont}[1]{S#1}
%%%%

\section{Derivation of the Hall conductivity}
\label{supp:DerivationHallConductivity}

Following Ref.~\citenum{supp_Szunyogh2009}, we summarize the derivation of the
Hall conductivity using linear-response theory. For a general, time-dependent
perturbation $H_{\mathrm{pert}}(t)$ acting on a system described by the
unperturbed, time-independent Hamiltonian $H_{0}$:
\begin{equation}
H(t) = H_{0} + H_{\mathrm{pert}}(t),
\end{equation}
the Hall conductivity tensor $\sigma_{\mu\nu}(\vect{q}, \omega)$ expresses the
linear response of a current in direction $\nu$ to an electric field applied in
direction $\mu \ne \nu$.  It can be expressed in terms of the current-current
correlation function $\Sigma_{\mu\nu}(\vect{q}, \omega)$:
\begin{align}
\label{sigmaxyLinearResponse}
\sigma_{\mu\nu}(\vect{q}, \omega) &= -\lim_{\eta \to 0^{+}}
\frac{\Sigma_{\mu\nu}(\vect{q}, \omega + i \eta) - \Sigma_{\mu\nu}(\vect{q},
0)}{\varpi}, \\
\Sigma_{\mu\nu}(\vect{q}, \omega) &= \frac{1}{\hbar V} \int_{0}^{\infty} dt\
e^{i \omega t} \trace \{ \rho_{0} [j_{\mu}(\vect{q}, t), j_{\nu}(-\vect{q}, 0)]
\},
\end{align}
with $\rho_{0} = |{0}\rangle \langle {0}|$ the zero-temperature density matrix
operator expressed in terms of the ground state $|{0}\rangle$ of the system.
In the long-wavelength and static limit ($\vect{q} \to 0$ and $\omega \to 0$)
we are interested in, the conductivity tensor simplifies to:
\begin{equation}
\label{eq:sigmaxyStaticLimit}
\sigma_{\mu\nu}(\vect{q} = 0, \omega = 0) = -\lim_{\eta \to 0^{+}}
\frac{\Sigma_{\mu\nu}(\vect{q} = 0, i \eta) - \Sigma_{\mu\nu}(\vect{q} = 0,
0)}{i \eta} = -d_{\varpi} \Sigma_{\mu\nu}(\vect{q} = 0, \varpi) \bigr|_{\varpi
= 0}
\end{equation}
with $\varpi \equiv \omega + i \eta$. In the following, we drop the momentum
argument, writing $\Sigma_{\mu\nu}(\vect{q} = 0, \varpi) \equiv
\Sigma_{\mu\nu}(\varpi)$ and $j_{\mu}(0, t) \equiv j_{\mu}( t)$.  We then
obtain:
\begin{subequations}
\begin{align}
\Sigma_{\mu\nu}(\varpi)
&= \frac{1}{\hbar V} \int_{0}^{\infty} dt\ e^{i \varpi t} \trace \{ |{0}\rangle
\langle{0}| (j_{\mu}(t) j_{\nu}(0) - j_{\nu}(0) j_{\mu}(t)) \} \\
&= \frac{1}{\hbar V} \int_{0}^{\infty} dt\ e^{i \varpi t} \langle{0}|
(j_{\mu}(t) j_{\nu}(0) - j_{\nu}(0) j_{\mu}(t))|{0}\rangle \\
&= \frac{1}{\hbar V} \int_{0}^{\infty} dt\ e^{i \varpi t} \sum_{n} (\langle{0}|
j_{\mu}(t) |{n}\rangle \langle{n}| j_{\nu}(0) |{0}\rangle - \langle{0}|
j_{\nu}(0) |{n}\rangle \langle{n}| j_{\mu}(t) |{0}\rangle) \\
&= \frac{1}{\hbar V} \int_{0}^{\infty} dt\ e^{i \varpi t} \sum_{n \neq 0}
(\langle{0}| j_{\mu}(t) |{n}\rangle \langle{n}| j_{\nu}(0) |{0}\rangle -
\langle{0}| j_{\nu}(0) |{n}\rangle \langle{n}| j_{\mu}(t) |{0}\rangle).
\end{align}
\end{subequations}
Evaluating the time dependencies of the Heisenberg operators, $A(t) = e^{i t
H_{0}/\hbar} A e^{-i t H_{0}/\hbar}$, then leads to:
\begin{equation}
\langle{0}| j_{\mu}(t) |{n}\rangle = \langle{0}| e^{i t H_{0}/\hbar} j_{\mu}
e^{-i t H_{0}/\hbar} |{n}\rangle = e^{i t E_{0}/\hbar} \langle{0}| j_{\mu}
|{n}\rangle e^{-i t E_{n}/\hbar} = e^{-i t \omega_{n0}} \langle{0}| j_{\mu}
|{n}\rangle,
\end{equation}
where $E_{n}$ is the energy of the $n$-th state, and $\omega_{n0} \equiv (E_{n}
- E_{0})/\hbar$.  For $\Sigma_{\mu\nu}(\varpi)$ we then obtain the following
relation:
\begin{subequations}
\label{eq:CurrentCurrentCorrelationFn}
\begin{align}
\Sigma_{\mu\nu}(\varpi)
&= \frac{1}{\hbar V} \int_{0}^{\infty} dt\ e^{i \varpi t} \sum_{n \neq 0}
\bigl( \langle{0}| j_{\mu} |{n}\rangle \langle{n}| j_{\nu} |{0}\rangle e^{-i t
\omega_{n0}} - \langle{0}| j_{\nu} |{n}\rangle \langle{n}| j_{\mu} |{0}\rangle
e^{i t \omega_{n0}} \bigr) \\
&= \frac{i}{\hbar V} \sum_{n \neq 0} \biggl( \frac{\langle{0}| j_{\mu}
|{n}\rangle \langle{n}| j_{\nu} |{0}\rangle}{\varpi - \omega_{n0}} -
\frac{\langle{0}| j_{\nu} |{n}\rangle \langle{n}| j_{\mu} |{0}\rangle}{\varpi +
\omega_{n0}} \biggr).
\end{align}
\end{subequations}
Substituting the result of Eq.~\eqref{eq:CurrentCurrentCorrelationFn} back into
Eq.~\eqref{sigmaxyLinearResponse} for $\vect{q} = 0$ we obtain:
\begin{subequations}
\label{eq:sigmaxyLinearResponse2}
\begin{align}
\sigma_{\mu\nu}(\omega)
&= -\lim_{\eta \to 0^{+}} \frac{\Sigma_{\mu\nu}(\varpi) -
\Sigma_{\mu\nu}(0)}{\varpi} \\
&= -\lim_{\eta \to 0^{+}} \frac{1}{\varpi} \frac{i}{\hbar V} \sum_{n \neq 0}
\biggr[ \langle{0}| j_{\mu} |{n}\rangle \langle{n}| j_{\nu} |{0}\rangle \biggl(
\frac{1}{\varpi - \omega_{n0}} - \frac{1}{-\omega_{n0}} \biggr) - \langle{0}|
j_{\nu} |{n}\rangle \langle{n}| j_{\mu} |{0}\rangle \biggl( \frac{1}{\varpi +
\omega_{n0}} - \frac{1}{\omega_{n0}} \biggr) \biggr] \\
&= -\lim_{\eta \to 0^{+}} \frac{i}{\hbar V} \sum_{n \neq 0} \biggl(
\frac{\langle{0}| j_{\mu} |{n}\rangle \langle{n}| j_{\nu} |{0}\rangle}{(\varpi
- \omega_{n0})\omega_{n0}} + \frac{\langle{0}| j_{\nu} |{n}\rangle \langle{n}|
j_{\mu} |{0}\rangle}{(\varpi + \omega_{n0})\omega_{n0}} \biggr) \\
&= -\frac{i}{\hbar V} \sum_{n \neq 0} \biggl( \frac{\langle{0}| j_{\mu}
|{n}\rangle \langle{n}| j_{\nu} |{0}\rangle}{(\omega - \omega_{n0})\omega_{n0}}
+ \frac{\langle{0}| j_{\nu} |{n}\rangle \langle{n}| j_{\mu}
|{0}\rangle}{(\omega + \omega_{n0})\omega_{n0}} \biggr).
\end{align}
\end{subequations}
In the static limit we find:
\begin{equation}
\label{eq:sigmaxyStaticLimit2}
\sigma_{\mu\nu}(\vect{q} = 0, \omega = 0) = \frac{i}{\hbar V} \sum_{n \neq 0}
\frac{\langle{0}| j_{\mu} |{n}\rangle \langle{n}| j_{\nu} |{0}\rangle -
\langle{0}| j_{\nu} |{n}\rangle \langle{n}| j_{\mu}
|{0}\rangle)}{\omega_{n0}^2}
\end{equation}
which after integration over the Brillouin zone BZ leads for $\mu = x$ and $\nu
= y$ to:
\begin{equation}
\label{eq:sigmaxyStaticLimit3}
\sigma_{xy} = - \frac{V}{(2\pi)^{2}} \int_{\mathrm{BZ}} d^{2}k \frac{2}{\hbar
V} \sum_{n \neq 0} \frac{\IM(\langle{0}| j_{x} |{n}\rangle \langle {n}| j_{y}
|{0}\rangle)}{(E_{n} - E_{0})^{2}/\hbar^2} = - \frac{1}{h \pi}
\int_{\mathrm{BZ}} d^{2}k \sum_{n \neq 0} \frac{\IM(\langle{0}| j_{x}
|{n}\rangle \langle {n}| j_{y} |{0}\rangle)}{(E_{n} - E_{0})^{2}/\hbar^2}.
\end{equation}
Rewriting the latter in terms of the Berry curvature (see
Ref.~\citenum{supp_Xiao2010}), and applying Stokes theorem to convert to an
expression in terms of the gauge-dependent Berry connection leads to:
\begin{equation}
\sigma_{xy} = -\frac{1}{h \pi} \int_{\mathrm{DP}} d^{2}k \sum_{n \neq 0}
\frac{\IM(\langle{0}| j_{x} |{n}\rangle \langle {n}| j_{y} |{0}\rangle)}{(E_{n}
- E_{0})^{2}/\hbar^2} = \frac{e^{2}}{2 \pi h} \int_{\mathrm{DP}} d^{2}k\
\Omega_{xy}(\vect{k}) = \frac{e^{2}}{2 \pi h} \oint_{\partial \mathrm{DP}}
d\vect{k} \cdot \vect{A}(\vect{k}).
\end{equation}
The integrals over momentum space are taken over the region around the Dirac
point DP, bounded by $\partial \mathrm{DP}$, over which the surface states lie
inside the bulk gap.

We now specialize to the case of the generalized Dirac model discussed in the
main text for which $\partial \mathrm{DP}$ is a circle with radius $\Lambda$
such that $v \Lambda \gg m$. Using the chain rule we find that
\begin{equation}
\sigma_{xy} = \frac{e^{2}}{2 \pi h} \oint_{\partial \mathrm{DP}} d\vect{k}
\cdot \vect{A}(\vect{k}) = \frac{e^{2} i }{2 \pi h}  \int_{0}^{2\pi} d\phi\
\langle{0}| \frac{\partial}{\partial \phi} |{0}\rangle\vert_{k=\Lambda},
\end{equation}
where $\phi$ is the orientation angle in momentum space.  This is
Eq.~\eqref{eq:sigmaxyOurModelA} of the main text.

As explained in the main text, to apply the Berry connection formula for the
Hall conductivity we must choose a gauge in which $|{0}\rangle$ is a smooth
function of momentum in $\mathrm{DP}$.

\subsection{Electron-spin-wave interaction vertex}
\label{supp:ElectronSpinWave}

\begin{figure}[tb]
\centering
\includegraphics[width=0.75\textwidth]{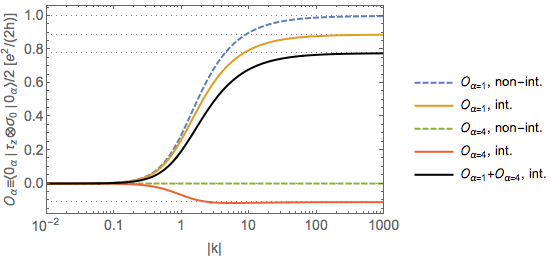}
\caption{%
Contribution to the expectation values $\mathcal{O}_{\alpha} \equiv
\langle{0_{\alpha}}| \tau_{z} \otimes \sigma_{0} |{0_{\alpha}}\rangle$ from
component $\alpha$ of the ground state wavefunction $|{0}\rangle$, as function
of momentum $|k|$. ($\tau_{z} \otimes \sigma_{0}$ is a diagonal operator with
diagonal components $(1, 0, 0, -1)$ so that only the first and fourth
components of the wavefunctions contribute.  In the non-interacting case
(dashed lines) the Berry phase comes only from $|{n=0}\rangle \otimes
|{v}|\rangle$, \textit{i.e.}\ the first component of the dressed ground state
wavefunction. In the spin-wave case (solid lines), at large $|k|$ the weight
and hence the Berry connection contribution from this component is reduced, and
a contribution of opposite sign from the $|{n=1}\rangle \otimes |{v}\rangle$
($\alpha = 4$) component arises.  The topological magneto-electric effect is
proportional to the sum of those Berry phases, $\mathcal{O}_{1} +
\mathcal{O}_{4}$ (black solid line).%
}
\label{fig:ExpectationValueTauzSigma0PerComponent}
\end{figure}

Due to the simple $\phi$ dependence of the perturbed ground state, explicitly
given in Eq.~\eqref{eq:pertgroundstate} of the main text, differentiating with
respect to $\phi$ is equivalent to simply multiplying $|{0}\rangle$  by the
diagonal matrix with non-zero entries $(-i, 0, 0, i)$.  This operator is
conveniently expressed as $-i (\tau_{0} \otimes \sigma_{z} + \tau_{z} \otimes
\sigma_{0})/2$ so that
\begin{equation}
\sigma_{xy} = \frac{e^{2}}{2 \pi h} \int_{0}^{2\pi} d\phi\, i \, \langle{0}|
\frac{\partial}{\partial \phi} |{0}\rangle\vert_{k=\Lambda} = \frac{e^{2}}{2 h}
\langle{0}| (\tau_{0} \otimes \sigma_{z} + \tau_{z} \otimes \sigma_{0})
|{0}\rangle \vert_{k=\Lambda},
\end{equation}
where $\tau_{\alpha}$ are Pauli matrices in the $n = 0, 1$ boson occupation
number space, and $\sigma_{\alpha}$ are Pauli matrices in spin space.  In the
limit $\Lambda \to \infty$ we find $\langle{0}| (\tau_{0} \otimes \sigma_{z})
|{0}\rangle \vert_{k=\Lambda} = 0$, because at large $|k|$ the expectation
value of $\sigma_{z}$ in each boson sector is zero.  Thus, $\sigma_{xy}$ is
equivalent to
\begin{equation}
\label{eq:sigmaxyOurModel3}
\sigma_{xy} = \frac{e^{2}}{2 h} \langle{0}| (\tau_{z} \otimes \sigma_{0})
|{0}\rangle \vert_{k=\Lambda}.
\end{equation}
The expectation value $\mathcal{O}_{\alpha} \equiv \langle{0_{\alpha}}|
\tau_{z} \otimes \sigma_{0} |{0_{\alpha}}\rangle$ of the individual components
of the ground state wavefunction is shown in
Fig.~\ref{fig:ExpectationValueTauzSigma0PerComponent}.

\subsection{General electron-boson interaction vertex}
\label{supp:GeneralInteractionVertex}

For an arbitrary interaction vertex $\mathcal{M}$, and arbitrary surface-state
band-structure model, we obtain the following perturbative expression to
leading order in $\Omega/\omega$:
\begin{equation}
\sigma_{xy} \approx \frac{e^{2}}{h} \biggl[ \nu + \biggl(
\frac{\Omega}{\omega} \biggr)^{2} \int_{\mathrm{DP}} \frac{d^{2}k}{4\pi}
f(\vect{d}(\vect{k}), \mathcal{M}) \biggr],
\end{equation}
where $\nu$ is given by:
\begin{equation}
\nu = \int_{\mathrm{DP}} \frac{d^{2}k}{4\pi}\ \vect{\hat{d}}(\vect{k}) \cdot
\bigl( \partial_{k_{x}} \vect{\hat{d}}(\vect{k}) \times \partial_{k_{y}}
\vect{\hat{d}}(\vect{k}) \bigr),
\end{equation}
and the function $f(\vect{d}(\vect{k}), \mathcal{M})$ is defined by:
\begin{multline}
f(\vect{d}(\vect{k}), \mathcal{M}) = \bigl[ \vect{\hat{d}}(\vect{k}) \cdot
\bigl( \partial_{k_{x}} \vect{\hat{d}}(\vect{k}) \times \partial_{k_{y}}
\vect{\hat{d}}(\vect{k}) \bigr) \bigr] \bigl\{ \hat{d}_{z}(\vect{k})
(|\mathcal{M}_{12}|^{2} - |\mathcal{M}_{21}|^{2}) \\
+ \hat{d}_{y}(\vect{k}) \bigl[ \RE(\mathcal{M}_{11} - \mathcal{M}_{22})
\IM(\mathcal{M}_{12} + \mathcal{M}_{21}) \bigr) - \IM(\mathcal{M}_{11} -
\mathcal{M}_{22}) \RE(\mathcal{M}_{12} + \mathcal{M}_{21}) \bigr] \\
- \hat{d}_{x}(\vect{k}) \bigl[ \RE(\mathcal{M}_{11} - \mathcal{M}_{22})
\RE(\mathcal{M}_{12} - \mathcal{M}_{21}) + \IM(\mathcal{M}_{11} -
\mathcal{M}_{22}) \IM(\mathcal{M}_{12} - \mathcal{M}_{21}) \bigr] \bigr\}.
\end{multline}
Due to the rotational symmetry of the Dirac model, the contributions
proportional to $\hat{d}_{x}$ and $\hat{d}_{y}$ vanish upon integration over
the Brillouin zone, since $\hat{d}_{x}(-\vect{k}) = -\hat{d}_{x}(\vect{k})$.
On the other hand, the contribution proportional to $\hat{d}_{z}$ yields a
finite correction since generally $\hat{d}_{z}(\vect{k})$ is an even function
of $\vect{k}$.  Therefore, for the Dirac model only off-diagonal terms in the
electron-boson interaction vertex $\mathcal{M}$ lead to a correction:
\begin{equation}
\sigma_{xy} \approx \frac{e^{2}}{2h} \SIGN{(\mathcal{V})} \biggl[ \SIGN (m) +
\frac{1}{2} \biggl(\frac{\Omega}{\omega} \biggr)^{2} (|\mathcal{M}_{12}|^{2} -
|\mathcal{M}_{21}|^{2}) \biggr].
\end{equation}
In particular, this implies that for an electron-phonon interaction vertex
described by $\mathcal{M}_{\mathrm{ph}} = \gamma_{\mathrm{ph}} \sigma_{0}$
there is no perturbation.  However, for the spin-wave interaction described by
$\mathcal{M}_{\mathrm{sw}} = \gamma_{\mathrm{sw}} (\sigma_{x} - i
\sigma_{y})/2$ we obtain a finite correction, as discussed in detail in the
main text.

%%%%%%%%%%%%%%%%%%%%%%%%%%%%%%%%%%%%%%%%%%%%%%%%%%%%%%%%%%%%%%%%%%%%%%%%%%%%%%%%

\section{Unchanged quantized Hall conductivity for static perturbations}
\label{supp:QuenchedSpinWave}

The absence of a disorder-induced correction to the Hall conductivity is
easiest to establish explicitly in the case of a spin-dependent but spatially
homogeneous perturbation, $\mathcal{H}_{\mathrm{pert}}^{\mathrm{st}} = g_{0}
\sigma_0 + \vect{g} \cdot \vect{\sigma}$, on top of the unperturbed
quasiparticle Hamiltonian, Eq.~\eqref{eq:BareElHamiltonian} of the main text.
Using the relationship between Berry phases and spin-coherent state
orientations for spin-$1/2$ we find that
\begin{equation}
\label{eq:ConstantPerturbation}
\sigma_{xy} = \frac{e^{2}}{4 \pi h} \int_{\mathrm{DP}} d^{2}k
\frac{\bigl(\vect{ d}(\vect{k}) + \vect g\bigr) \cdot \bigl( \partial_{k_{x}}
\vect{ d}(\vect{k})\times \partial_{k_{y}} \vect{d}(\vect{k})
\bigr)}{|\vect{d}(\vect{k}) + \vect{g}|^{3}}.
\end{equation}
Performing the integration in Eq.~\eqref{eq:ConstantPerturbation} for the
massive Dirac Hamiltonian model (Eq.~\eqref{eq:MassiveDiracHamiltonian} in the
main text) we find that at zero temperature
\begin{equation}
\sigma_{xy} = \SIGN(\hbar m + g_{z}) \frac{e^{2}}{2h}.
\end{equation}
Therefore, there is no change in the Hall conductivity unless the perturbation
is sufficiently large to change the gap. The corresponding derivation for
finite wavevector $\vect{p}$ spatially modulated perturbations follows exactly
the same lines and, because the current operator of the Dirac model is
independent of wavevector, leads to an expression that is identical to
Eq.~\eqref{eq:ConstantPerturbation}.

%%%%%%%%%%%%%%%%%%%%%%%%%%%%%%%%%%%%%%%%%%%%%%%%%%%%%%%%%%%%%%%%%%%%%%%%%%%%%%%%

\end{document}